\documentclass[]{jfm}

\usepackage{graphicx}
\usepackage{newtxtext}
\usepackage{newtxmath}
\usepackage{natbib}
\usepackage{hyperref}
\hypersetup{
    colorlinks = true,
    urlcolor   = blue,
    citecolor  = black,
}

\usepackage{siunitx}

\newcommand{\RomanNumeralCaps}[1]

\title{Spreading of a viscoelastic drop on a solid substrate}

\author{{Peyman Rostami\aff{1}
  \corresp{\email{rostami@ipfdd.de}},
Mathis Fricke \aff{2},
Simon Schubotz \aff{1,3},
Himanshu Patel\aff{1},
Reza Azizmalayeri \aff{1},  
G\"unter K. Auernhammer \corresp{\email{auernhammer@ipfdd.de}}}}

\affiliation{\aff{1} Leibniz-Institut f\"ur Polymerforschung Dresden e.V.,Dresden, Germany,
\aff{2}Department of Mathematics, TU Darmstadt, Darmstadt, Germany,
\aff{3}Technische Universität Dresden, Dresden, Germany}

\begin{document}
\maketitle

\begin{abstract}
We study the spreading of viscous and viscoelastic drops on solid substrates with different wettability. In the early stages of spreading, we find  that the viscoelastic drop spreads faster and has a different power law than the Newtonian drop (i.e. aqueous glycerine solution) for the same zero shear rate viscosity. We argue that the effect of viscoelasticity is only observable for  experimental time scales in the order of the internal relaxation time of the polymer solution or longer times. Near the contact line, the effective viscosity is lower for the viscoelastic drop than for the Newtonian drop. Together with its shear rate dependency, this difference in effective viscosity can explain the different spreading dynamics. 
We support our experimental findings with a simple perturbation model that qualitatively agrees with our findings.  

\end{abstract}

\begin{keywords}
Drop spreading, Viscoelastic liquids
\end{keywords}

 \section{Introduction}
\label{sec:introduction}

For at least the last two centuries, the interaction of droplets with surfaces has been studied quantitatively. 
In retrospective outstanding work was done by Young for static wetting (\cite{young1805iii}), Worthington for drop impact (\cite{Worthington-Arthur:1883aa}) and drop spreading over different surfaces (\cite{hardy1919iii, shuttleworth1948spreading, fox1950spreading}).
Drop spreading and its dynamics play an essential role in many industrial applications, from printing to coating (\cite{hoath2016fundamentals,Sankaran:2012aa, glasser2019tuning}).
The spreading of Newtonian drops has been the subject of an extensive research over the last two decades (\cite{biance2004first, bonn2009wetting,bird2008short,snoeijer2013moving}).
For low-viscosity drops, the key finding is that the spreading dynamics consist of two regimes; an inertial and a viscous dominated regime (\cite{biance2004first}).

The boundary condition has an important influence on the calculation of the flow field close to the contact line and thus on the viscous dissipation.
By assuming a non-slip condition, the contact line motion is solved by \cite{Moffatt:1964aa}.
The assumption of non-slip condition leads to a divergence of the viscous stress due to the hydrodynamic singularity at the moving contact line (\cite{Huh:1971aa, tanner1979spreading,huh1971hydrodynamic,huh1977steady}). 
Various solutions  have been proposed to this problem in molecular scale (\cite{BLAKE:1969aa}), hydrodynamic models by (\cite{COX:1986aa},\cite{Voinov:1976ab} and \cite{SHIKHMURZAEV:1997aa, shikhmurzaev2020moving}) and by including the evaporation (\cite{Rednikov:2012aa}) .
For the fast processes, the dynamics can be modeled by the hydrodynamic model with a slip length that generates a  lower cut-off length below which the liquid and solid velocities are allowed to differ in the vicinity of the contact line and/or the substrate. 

Consider a drop of an initial radius  $R$, volume of $V$, density  $\rho$, viscosity  $\eta$, and surface tension  $\sigma$ which gently touches a solid substrate, it starts spreading with velocity of $u$. 
In the early stage of spreading, inertia is assumed to be dominant (\cite{biance2004first}). 
By writing a force balance between inertial $\frac{d }{d t}((\rho V)u)$ and capillary forces  $\sim \sigma r$, one can derive the spreading rate, Eq.~(\ref{Inertia}), where the radius of the wetted on the substrate area is $r$.

\begin{equation}\label{Inertia}
\left(\frac{r}{R}\right)^{2}= t\sqrt{\frac{\sigma R^{3}}{\rho }}
\end{equation}

In a second regime, the viscous dissipation near the contact line is the rate-limiting process when the drop shape is close to a spherical cap. 
For the viscous dominated  regime (Tanner regime) of drop spreading, the Cox-Voinov relation for the dynamic contact angle in case of perfect wetting is $\theta^{3}\sim \eta \frac{u}{\sigma }$, and the conservation of volume, $r^{3}\theta \sim V$, results in the spreading dynamic in the viscous regime. 
This relation is known as Tanner's law $r \sim  R \left(\frac{\sigma t}{\eta R}\right)^{\tfrac{1}{10}}$(\cite{tanner1979spreading}).
It should be noted that Cox-Voinov was originally developed for a final contact angle $\theta = 0$, but was later shown to be valid for higher contact angles up to \SI{100}{\degree} (\cite{fermigier1991experimental, petrov2003dynamics}).
By equating the radius from the inertial and viscous regimes, the transition between these two regimes can be calculated: $\tau_{iv} \sim (\frac{\rho \sigma R}{\eta ^{2}})^\frac{1}{8}\sqrt{\frac{\rho R^{3}}{\sigma }}$ (\cite{biance2004first}).

The above models work reasonable for low viscosity drops (e.g. water). 
For the early stages of high-viscosity drop spreading, there are several conflicting results (\cite{carlson2011measuring,carlson2012universality,eddi2013short}).
\cite{carlson2012universality} stated that the drop spreading dynamics still follow the power-law type of spreading with same exponent but with a friction factor ($\mu _{f}$) as a correction factor for the prefactor of the power law, $\frac{r}{R}\sim (\frac{\sigma t}{R \mu_{f}})^{\frac{1}{2}}$. 
\cite{eddi2013short} argue that, for high viscous liquids, the inviscid solution is not valid anymore, so they solve Stokes flow in this case. 
An important approach is to use the assumed analogy between the merging of identical drops (\cite{Eggers:1999aa}) and the spreading of drops on a substrate.
\cite{eddi2013short} used a logarithmic model to scale their experimental data $r\simeq -\frac{1}{4\pi }\frac{\sigma }{\eta }t \ln\left(\frac{t}{R}\right)$ .

Despite many industrial applications (e.g. printing), the early drop spreading of viscoelastic fluids has not been extensively studied.
In most of the studies, the viscous dominated drop spreading (late stage) is studied experimentally and numerically (\cite{carre2000spreading,betelu2003capillarity,liang2009spreading,iwamatsu2017spreading} \cite{jalaal2021spreading, iwamatsu2017spreading}).
The viscous spreading exponent ($\alpha$) is correlated with the rheological exponent $n$ in these models.
Just very recently, the early stage of drop spreading of shear thinning fluids is studied by two groups (\cite{yada2023rapid,bouillant2022rapid}).
Both groups reported that the early stage of drop spreading (regardless of polymer concentration and molar mass) follows the same trend as low viscosity drop spreading (e.g. water drops).
The considered time scale of experiments, in both cases are bellow inertia-capillary  time ($\tau_{ic}=\sqrt{\frac{\rho R^{3}}{\sigma}}$) which is in the order of few milliseconds for millimetric drops. 

Viscoelastic materials combine an elastic component and a viscous component  in their properties. 
When a polymer solution is stretched, in the beginning only the elastic part  contributes to the dynamics and  after a  characteristic time the viscous part part becomes relevant (\cite{costanzo2016shear}). 
To observer the effect of viscoelasticity, the experimental time scale should be in order of viscoelastic time scale, i.e. polymer relaxation time.  
In a simple approach, the viscosity of polymer solutions can be described by the Cross-model (\cite{gastone2014green,subbaraman1971extrapolation,cross1965rheology}).
It is shown that when shear is applied to a viscoelastic material, it takes several times of the relaxation time of the sub chain to reach a steady state.
The relaxation time depends on the polymer concentration or/and molar mass (number of entanglements) (\cite{costanzo2016shear, VEREROUDAKIS2023105021}).

\begin{equation}\label{cross-fluid-model}
\eta =\frac{\eta _{0}-\eta _{\infty }}{1+(\tau_{ve} \dot{\gamma })^{m}}+\eta _{\infty }
\end{equation}

Here $\dot\gamma$ is the shear rate, $\tau_{ve}$ and $m$ are fluid parameters (polymer relaxation time and rheological exponent), and $\eta _{0}$ and $\eta _{\infty }$ are zero and infinite shear rate viscosities respectively. 
By increasing the polymer concentration and/or polymer molar mass, the polymer relaxation time $\tau_{ve}$ increases and the rheological exponent $m$ decreases (see SI). 

In this contribution, we follow the hypothesis that three time scales should be considered; the inertia-viscous ($\tau_{iv}$) cross over time, the inertia-capillary  cross over time ($\tau_{ic}$) and polymer relaxation time ($\tau_{ve}$). 
To illustrate these time scales, we calculate them for water and an aqueous PEO solution ($4\% (w/w)$ ) of the molar mass of ($6\times 10^{5} (\frac{g}{mol})$).
For a millimetric water drop we get, $\tau _{iv}\sim \SI{15}{\milli \second}$, $\tau _{ic}\sim \SI{3.7}{\milli \second}$ and $\tau_{ve} \sim 0$, and for the polymer solution $\tau _{iv}\sim \SI{2.8}{\milli \second}$, $\tau _{ic}\sim \SI{4}{\milli \second}$ and $\tau_{ve} \sim \SI{22}{\milli \second}$.    
In this contribution, we study some effects of these changes in the order of the time scales and provide a simple model to rationalize our findings.

\section {Experimental method}

\begin{figure}
\centering
\includegraphics[width=0.5\linewidth]{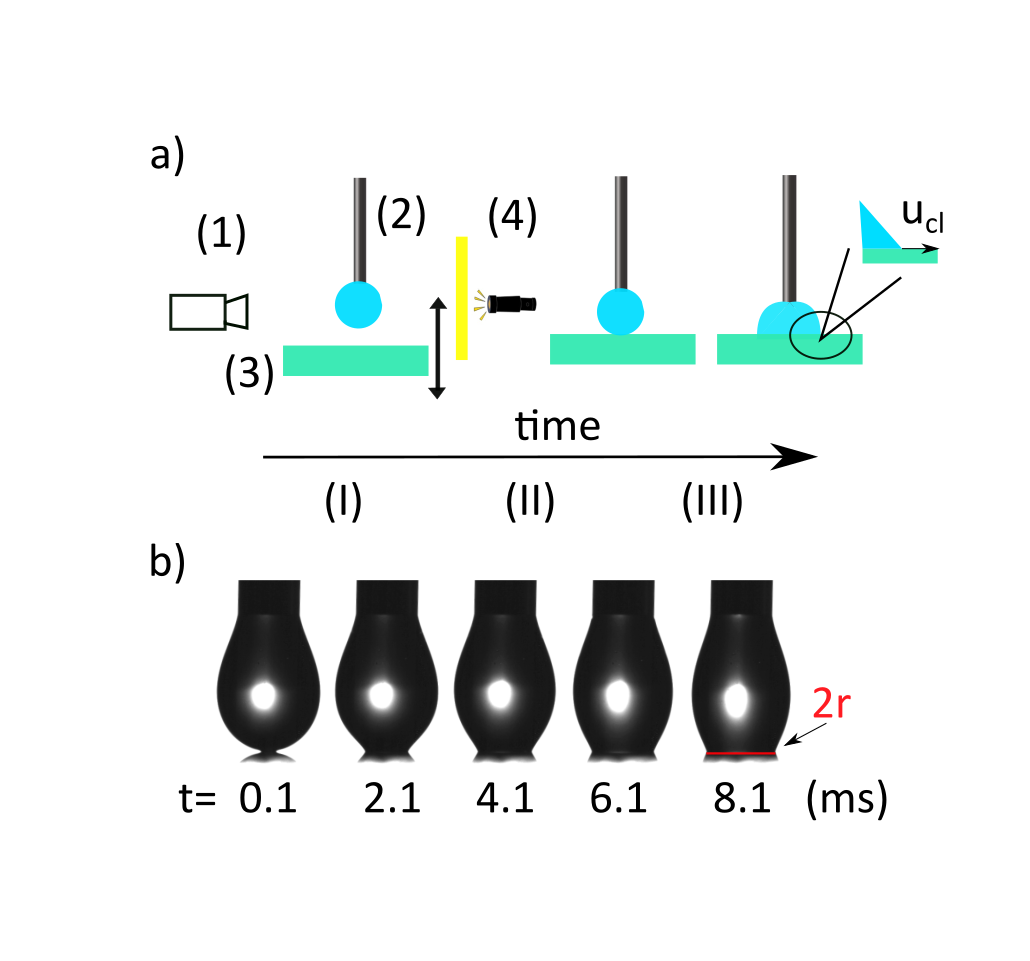}
\caption{a) Sketch of the  drop spreading setup, with high speed camera (1), light source and diffuser sheet (4), the drop and needle (2) and the adjustable solid substrate (3). Different stages of spreading are illustrated over time, (I) before contact between substrate and the drop (II) the substrate is  gently coming up and the drop touches the substrate (the initial point of contact). (III) After contact and spreading of drop on the substrate with contact line velocity of $u_{cl}$. b) The spreading dynamics of a milimetric drop over hydrophilic substrate for different times, the spreading radius $r$ is shown.}
\label{setup}
\end{figure}

The droplet dispenser is set up so that the drop hangs from a needle and the substrate is lifted up gently touching the drop.
After contact, the drop spreads immediately. 
The process is recorded in side view by a high-speed camera (FASTCAM Mini AX 200, Photron). 
The test section is illuminated by an LED lamp (SCHOTT KL 2500) along with a diffuser sheet to have homogeneous light in the background (Fig.~\ref{setup}).

Water (MicroPure UV/UF, Thermo Scientific Co.), glycerin (Sigma Aldrich co. 99\%), and mixtures thereof as well as water-polyethylene oxide (PEO, Sigma Aldrich co.) solutions of various molar masses and concentrations are used as Newtonian and viscoelastic operating fluids, respectively.
For aqueous PEO solution, from now on, the weight concentrations are shown as \% and the molar mass of polymers are mentioned as $k$ which represent $10^{3} \frac{g}{mol}$ .
The sample names and viscosities at zero shear rate (zero shear viscosity) for each liquid are given in  table~\ref{Liquids properties}.
The surface tension $\sigma$ of all samples is in the range of $ \SI{63}{\milli\newton\per\meter} \le \sigma \le \SI{72}{\milli\newton\per\meter}$. 
To measure the flow curves, a commercial rheometer (MRC 502, Anton-Paar GmbH) is used. 
For all measurements a cone-plate geometry is used with diameter of 50 ($mm$) and cone angle of $1^{\circ}$ and the gap of $100\mu m$ (CP50-1).
In the rheological experiments the temperature of the sample was kept constant  at \SI{20 \pm 1}{\celsius}. 
The rheological properties of each sample are given in the supplementary information.
Two types of surface coatings are used to study the effect of contact angle. 
Cleaned glass substrates as hydrophilic (contact angle of water drop around \SI{15}{\degree}) and silanised glass substrates as hydrophobic substrates  (contact angle of water drop around \SI{90}{\degree}) are used.
Details of substrate preparation are given in the supplementary information.
When preparing polymer solutions, it is crucial to wait long enough for the polymer to dissolve homogeneously in the solution.
This is illustrated by our rheology experiments.
For high molar masses, we measured changes in the flow curves within the first month after preparing the sample (see supplementary information).

\begin{table}
\centering
\begin{tabular}{lclll}
Sample  & Molar mass ($10^{3} \frac{g}{mol}$) & $\eta_{0}$ (\SI{}{\milli\pascal\cdot\second})   & Sample  & $\eta_{0}$ (\SI{}{\milli\pascal\cdot\second})  \\
Water + PEO ($2\%, 300k$) &$300$ & $35\pm 0.5$  & Water + Glycerin (0\%) & $0.93\pm 0.01$  \\
Water + PEO ($3\%, 300k$) &$300$  & $101\pm 0.5$  & Water + Glycerin (72\%) & $35\pm 0.5$ \\
Water + PEO ($4\%, 300k$) &$300$ & $254\pm 0.5$  & Water + Glycerin (85\%) & $98\pm 0.5$  \\
Water + PEO ($2\%, 600k$) &$600$  & $103\pm 0.5$  & Water + Glycerin (91.5\%)& $293\pm 0.5$  \\
Water + PEO ($3\%, 600k$) &$600$  &$537\pm 0.5$   & Water + Glycerin (93.5\%) &$389\pm 0.5$  \\
Water + PEO ($4\%, 600k$) &$600$  & $1324\pm 0.5$  &Water + Glycerin (100\%) & $1078\pm 0.5$ \\
\end{tabular}
\caption{\label{Liquids properties}Composition of operating fluids and the zero-shear viscosity $\eta_{0}$  (at \SI{23}{\celsius}).
The rheological properties of each sample are given in the supplementary information.} \end{table}

\section {Spreading of viscoelastic and viscous Newtonian drops }
 \subsection{Hydrophilic substrates}
 \begin{figure}
\centering
\includegraphics[width=0.75\linewidth]{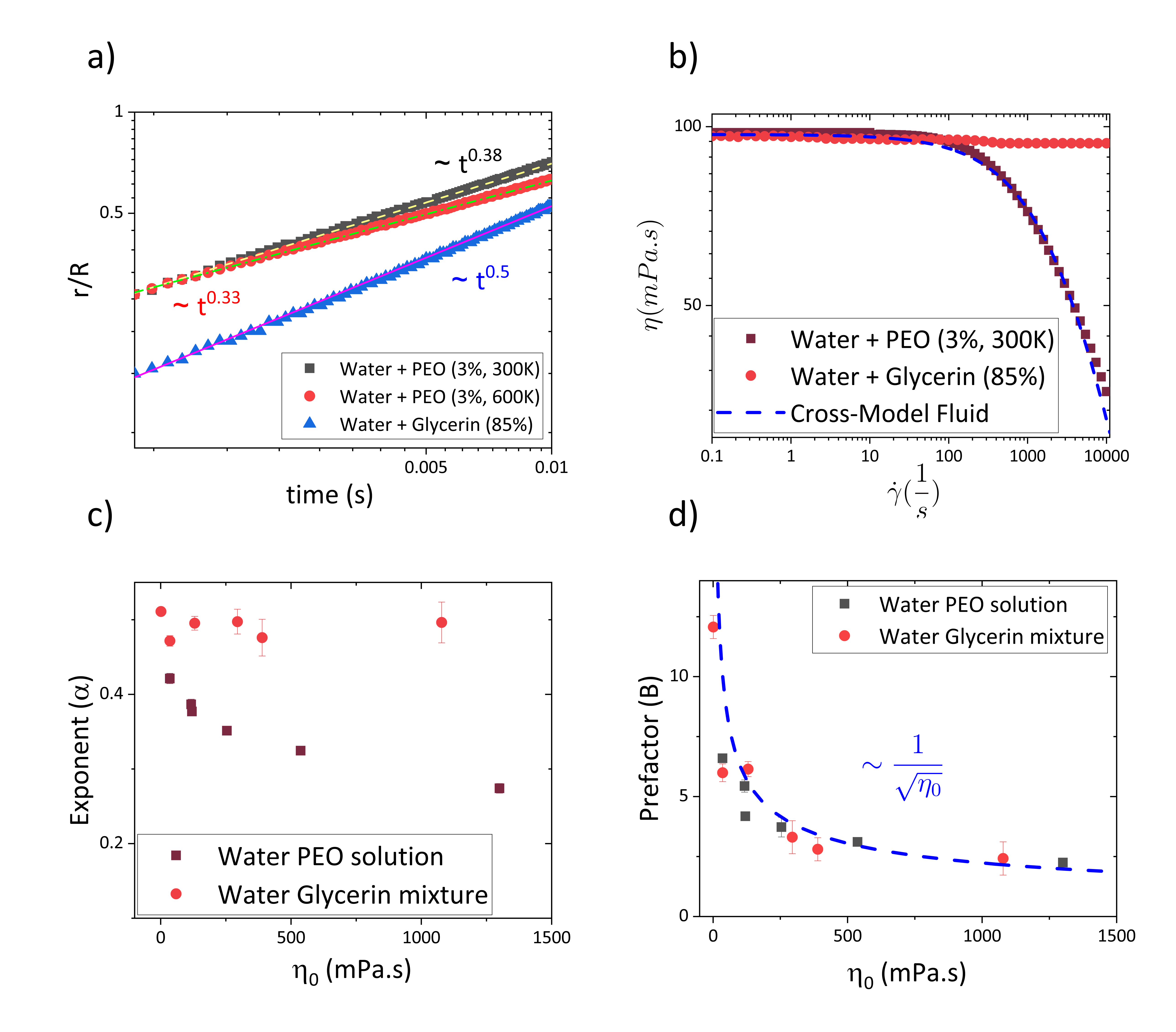}
\caption{a) Radius of wetted area ($r$) normalized by initial radius of drop ($R$) as a function of time, for water and PEO (3\%, 300k and 3\%, 600k) and water and glycerin (85\%) on hydrophilic substrates. b) Flow curve for water and PEO (3\%, 300k) solution and mixture of water and glycerin (85\%). c) and d) Spreading exponent ($\alpha$) and spreading prefactor as a function of zero-shear viscosity $\eta_{0}$ for all viscous Newtonian and viscoelastic liquids  mentioned in table~\ref{Liquids properties}.}
\label{spreading-hydrophilic}
\end{figure}

In Fig.~\ref{spreading-hydrophilic}a), we plot the radius of the wetted area ($r$) normalized to the initial drop radius ($R$), the initial spreading of viscous Newtonian (water-glycerol mixture) and viscoelastic drops on a hydrophilic substrate are plotted, (Fig.~\ref{spreading-hydrophilic}a). 
One of the experimental challenges of plotting the spreading radius over time is to determine the time of the first contact.
Three options have been suggested to overcome this problem, the simplest of which is to add a bottom view camera and capture the process from bottom (\cite{eddi2013short}).
Plotting the contact line velocity against the dimensionless spreading radius ($r/R$) is another possible approach (\cite{hartmann2021breakup}).
The advantage of this method is that it does not need the definition of zero time (see supplementary information).  
The third option is to define a fitting parameter as $t_{0}$, this parameter can pop up in the fitting function $r=B(t-t_{0})^{\alpha }$. 
In all our experiments, this parameter is of the order of a few frames $t_{0}\sim 0.0001 s$. 
In all plots the first four data points are omitted to be sure that the definition of the first contact time has no influence on the fit.
Consequently, we fitted the drop spreading results from the fifth data point to the polymer relaxation time scale ($\tau _{ve}$).

The viscous Newtonian drop spreads like  expected from Eq.~(\ref{Inertia}) proportional to square root of time, $r \sim \sqrt{t}$ however, aqueous PEO solutions spread with different spreading exponents ($ r = B \; t^\alpha$).
To illustrate this difference two samples with same zero-shear viscosities, here ''Water + PEO (3\%, 300k)'' and ''Water + Glycerin (85\%)'' ($\eta_0\approx \SI{100}{\milli\pascal\second}$), same initial drop size ($D \approx \SI{4}{\milli\meter}$), density, surface tension ($\sigma \approx \SI{65}{\milli\newton\per\meter}$) have clearly different  spreading exponents $\alpha$ (Fig.~\ref{spreading-hydrophilic}a). 
Based on the present models, these drops should spread in a same manner, the major difference between two samples is the viscoelasticity of second drop.
For the same prefactor in the power law, a smaller spreading exponent ($\alpha$) (for $t\leq \SI{0.01}{\second}$) results in higher spreading rates.

To illustrate the effect of viscoelasticity, we use a typical flow curve of PEO solutions. 
The viscosity at low shear rates remains constant.   
Above a certain critical shear rate, the viscosity decays with increasing the shear rate. 
Such a behavior can be described by the cross-fluid model, Eq.~(\ref{cross-fluid-model}). 
In contrast, water-glycerin mixtures show no evidence of shear thinning in our data, Fig.~\ref{spreading-hydrophilic}b. 
However, this may occur at even higher shear rates (\cite{Dontula:1999vl}).
The shear rate at which the viscosity differs from its zero shear value defines an internal relaxation time of the polymer solution, $\tau _{ve}$ in Eq.~(\ref{cross-fluid-model}).

Our hypothesis is that the spreading dynamics depend on the rheological properties of operating fluid.
The velocity of the contact line in this early regime of drop spreading is in the order of (\si{\meter\per\second}).
At \SI{1}{\micro\meter} from the contact line, the shear rate is estimated to be in the order of $10^6\:\si{\per\second}$. 
Most of the viscous dissipation occurs near the contact line (\cite{bodziony2023stressful}).
At these high shear rates, the viscosity of polymer solutions decreases significantly.
On the other hand, the viscosity of a Newtonian liquid remains more or less constant.
This means that in the Newtonian case, the effective viscosity near the contact line is higher. 
Consequently, the dissipation is higher in this case, leading to a lower contact line velocity, which is in good agreement with our experimental results.
This argument relies on the steady state viscosity of the liquid, i.e., when the polymer chains adapted to the shear rate and contribute to the dissipation in the polymer solution. 
For shorter times, the polymers cannot fully contribute to the dissipation in the polymer solution; the solution is not yet viscoelastic \cite{costanzo2016shear, VEREROUDAKIS2023105021}.

To test our hypothesis, we measure the spreading exponent and the spreading prefactor over a wide range of zero-shear viscosities and rheological properties  for viscous Newtonian and viscoelastic drops spreading on hydrophilic substrates, (Fig.~\ref{spreading-hydrophilic}c and d). 
The initial observations are verified by this systematic variation of the material parameters.
For viscous Newtonian fluids, the spreading exponent is slightly decreasing with increasing  zero-shear viscosity, $\alpha \approx 0.5$ on the hydrophilic substrates. 
In contrast, for viscoelastic fluids, the spreading exponent is a strongly decreasing function of the zero-shear viscosity, (Fig.~\ref{spreading-hydrophilic}c). 
This difference is an indication of the dependence of the spreading exponent ($\alpha$) on the rheological properties.
The trend of the spreading prefactor seems to depend only on the zero-shear viscosity but not on the rheological exponent (Fig.~\ref{spreading-hydrophilic}d). 
The spreading prefactor (for Newtonian and viscoelastic cases) roughly follows $B \sim {\eta_{0}}^{-0.5}$ Fig.~\ref{spreading-hydrophilic}d), which was also observed previously (\cite{carlson2011measuring, eddi2013short}).

\subsection{Effect of substrate's wettability}

Repeating the spreading experiments for Newtonian and viscoelastic liquids on hydrophobic substrates (i.e. $\theta_{0}\simeq 90^\circ$) reveals a number of important observations, Fig.~\ref{master-exponent} a.
(i) On average, the spreading exponent ($\alpha$) decreases as the contact angle increases.
(ii) The spreading exponent remains almost independent of zero-shear viscosity for Newtonian drops on hydrophobic substrates. 
(iii) Increasing the zero shear viscosity of viscoelastic liquids (i.e. increasing the concentration and/or molar mass of the polymer) reduces the exponent. 
The difference between the spreading of  Newtonian and viscoelastic drops shows the same trend regardless of the hydrophobicity of the substrate.
To summarize our experimental results, the spreading exponent is dependent on the viscoelasticity of drop and the prefactor is a function of viscosity. 
All of the developed models up to now cannot predict the effect of viscoelasticity, in the next section we present a simplistic model to predict this behavior.

\section{Modeling}
\subsection{Inviscid case}

For low viscosity drops, increasing the hydrophobicity of the substrate (by suitable surface modification of the substrate),  results in a decreasing spreading rate and exponent (\cite{bird2008short, chen2013initial, DU}).
This was explained in terms of a simple energy balance. 
This balance assumes that no energy is dissipated. 
In this approximation, the kinetic energy (left hand side of Eq.~\ref{Dissipationlowviscosity}) is balanced by a combination of free surface energy and wetting energy (right hand side of Eq.~\ref{Dissipationlowviscosity}, \cite{bird2008short}).
\begin{equation}\label{Dissipationlowviscosity}
\int_{V}\frac{1}{2}\rho u^{2}\text{d}V=\sigma\left[A(0)-A(t)+\pi r(t)^{2} \cos(\theta_{0})\right].
\end{equation}
Here $u$ is the velocity field inside the drop, $\rho$ is the liquid density, $A(t)$ and $A(0)$ are the surface area of the liquid-vapor interface during the spreading and at the time zero and $\theta_{0}$ is the contact angle at which the drop spreading would stop, i.e., the static advancing contact angle. 
\cite{bird2008short} solved the balance equation (by modeling the kinetic energy integral) and showed that the spreading is  a function of the substrate's wettability (Eq.~\ref{spreading-f(contactangel)}).   
\begin{equation}\label{spreading-f(contactangel)}
r(t)=c_{1}t^{\alpha}.
\end{equation}
In this solution, the spreading exponent is $\alpha=c_{2}\sqrt{F(\theta_{0})+\cos(\theta_{0})}$, where the unknown function $F$ depends weakly on $\theta_{0}$ (see \cite{bird2008short}). 
This means that as the contact angle increases, the spreading exponent decreases. 
Our experiments show the same behavior (Fig.~\ref{master-exponent} a). 

\subsection{Including viscous dissipation}

For viscous drops, viscous dissipation cannot be neglected.
The \cite{Moffatt:1964aa} solution suggests that the dissipation near the contact line is in the order of $\sim 2 \pi \eta r u^{2}$.
The dissipation rate is balanced by the rate of change of total energy.
We add the dissipation term to the left hand side of the time derivative of Eq.~\ref{Dissipationlowviscosity}. 
To solve the resulting equation (Eq.~\ref{perturbation-ODE}), we assume the viscous term to be small ($\eta \rightarrow \epsilon \eta^\ast $) and act as a perturbation term.
From now on we drop all explicit mentioning of time as an argument.

\begin{equation}
 2 \pi \sigma \left\{\frac{1}{2} t\left[\dot r\right]^{2}+\frac{1}{2} t^2 \left[\ddot r\right]-r\;\dot r\left[F(\theta _{0})+\cos(\theta_{0})\right]\right\}=-2 \pi \epsilon \eta^\ast r \dot r^2 
 \label{perturbation-ODE}
\end{equation}

We rewrite Eq.~\ref{perturbation-ODE} in dimensionless units ($t = \tau_{ve}\; \hat t $, $r = r^\ast\; \hat r $).
After simplification can be rewritten as Eq.~\ref{perturbation-ODE-dimless}.
Here the elastocapillary number is the characteristic dimensionless number $\mathrm{Ec}=\frac{\sigma \tau_{ve}}{\eta r}$. 
The elastocapillary number is the ratio between the capillary time scale and the polymer relaxation time scale. In our perturbation approach, it is convenient to have $\mathrm{Ec}$ on the right side.
We actually use the inverse of $\mathrm{Ec}$, as $\mathrm{Ec}^\ast = \frac{\eta^\ast r^\ast}{ \sigma \tau_{ve}}=\frac{1}{ \mathrm{Ec}}$.
Taking typical values for viscosity of the liquids ($\eta^\ast \rightarrow \SI{100}{\milli \pascal \second}$), the length scale as initial drop radius ($r^\ast \rightarrow \SI{2}{\milli \meter}$) and the time scale as the highest relaxation time of polymer solution ($\tau_{ve} \rightarrow \SI{22}{\milli \second}$), we get $\mathrm{Ec}^\ast \approx 0.12$.  

\begin{equation}
 \left\{\frac{1}{2} t\; \dot r^{2} +\frac{1}{2} t^2\; \ddot r -r\;\dot r \left[F(\theta _{0})+\cos(\theta_{0})\right]\right\}=-\epsilon \; \mathrm{Ec}^\ast r \; \dot r^2 
 \label{perturbation-ODE-dimless}
\end{equation}

\begin{figure}
\centering
\includegraphics[width=0.75\linewidth]{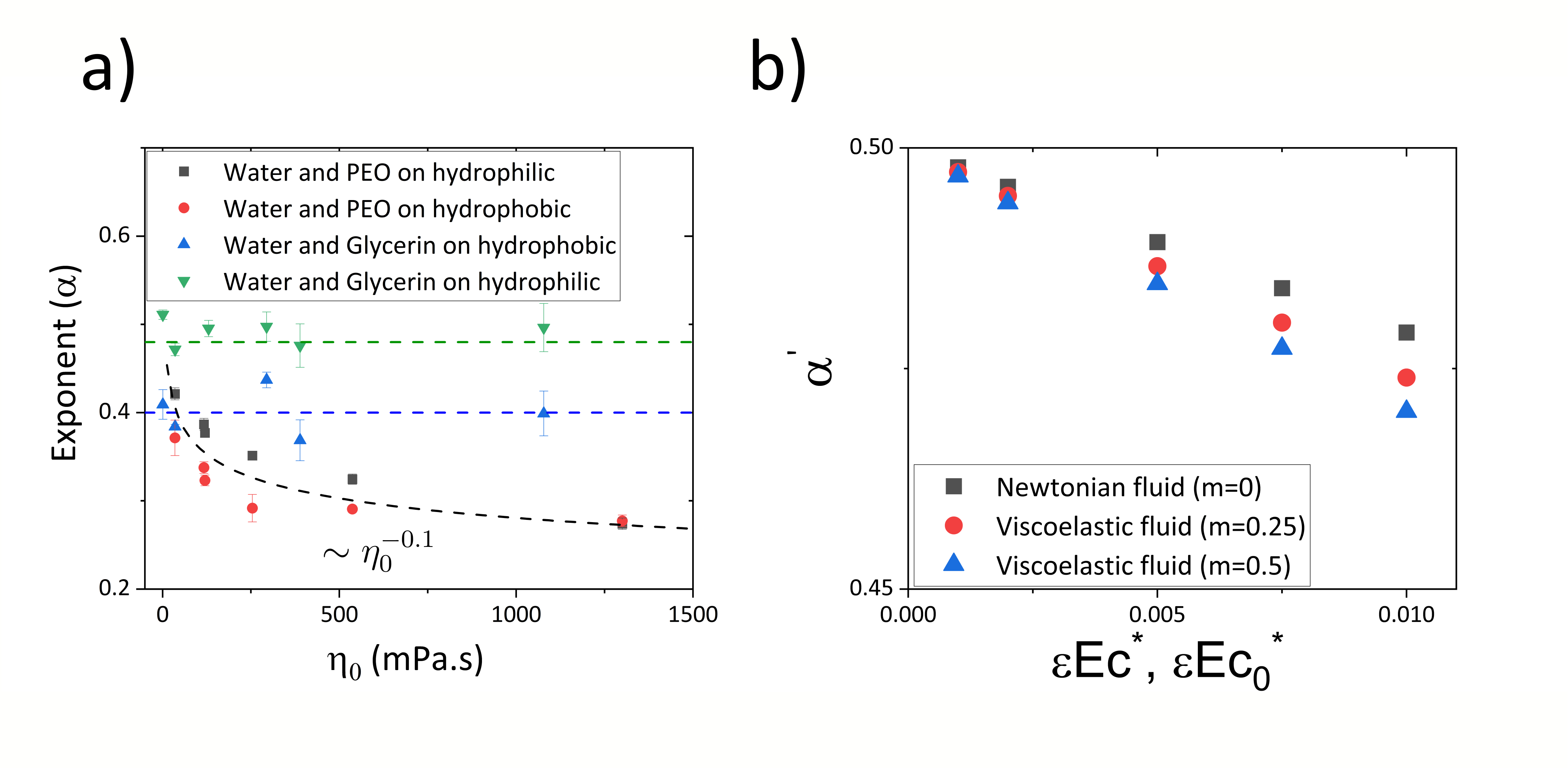}
\caption{ a) Experimental spreading exponent ($\alpha$) as a function of the zero-shear viscosity $\eta_{0}$ for viscous Newtonian and viscoelastic liquids  on hydrophobic and hydrophilic substrates. b)The theoretically predicted effective exponent($\alpha^{'}$) versus the different effective viscosity $\epsilon \mathrm{Ec}^{\ast }$ values, for viscous Newtonian and viscoelastic fluids ($m=0.25$ and  $m=0.5$).   }
\label{master-exponent}
\end{figure}

To address the viscoelastic case, we express the viscosity in Eq.~\ref{perturbation-ODE} as $\eta \rightarrow   \frac{\eta_{0}}{\dot{\gamma}^{m}} \rightarrow  \epsilon \frac{\eta^\ast_0}{\dot{\gamma}^{m}} $.
Note that  $\frac{\eta^\ast_0}{\dot{\gamma}^{m}}$ has the dimensions of a viscosity. 
Since mainly the high shear region close to the contact line contributes to the viscous dissipation, only the high shear-rate viscosity is considered here. 
By estimating the shear rate as a function of the contact line velocity as $\dot{\gamma} \simeq \frac{u_{cl}}{d^\ast}$ ($d^\ast$ is the distance to the contact line), the viscous dissipation can be written as $\simeq \eta^\ast ru^{2-m}$.
With this assumption, the force balance equation (Eq.~\ref{perturbation-ODE}) can be rewritten:
\begin{equation}
 \left\{\frac{1}{2} t\; \dot r^{2} +\frac{1}{2} t^2\; \ddot r -r\;\dot r \left[F(\theta _{0})+\cos(\theta_{0})\right]\right\}=-\epsilon \; \mathrm{Ec}^\ast_0 r \; \dot r^{2-m} 
 \label{perturbation-ODE-N-N-dimless}
\end{equation}
We use Mathematica (Wolfram Alpha Co. Version 10) to solve Eq.~\ref{perturbation-ODE-dimless} and \ref{perturbation-ODE-N-N-dimless} numerically, see SI for details.
We start exploring the effect of Newtonian and viscoelastic viscosity from the inviscid case discussed in  (\cite{bird2008short}), for example we consider the case that the drop spreading follows: $r(t)= 0.02t^{0.5}$, , in our dimensionless units. 
The obtained solution is not exactly a power law, but close to it.
We fitted the numerical results by a simple power law ($B^{'}t^{\alpha^{'}}$), where $B^{'}$ and $\alpha^{'}$ are the effective prefactor and exponent, respectively (see SI).
The theoretical exponents are plotted in Fig.~\ref{master-exponent} b as a function of $\epsilon$ and $m$. 
We consider the elastocapillary numbers are equal in both cases, since in experimental part we compare the drops with same zero shear viscosity .
These results show that in viscoelastic case, the exponent decreases stronger than in the Newtonian case, by increasing the $m$ value (i.e. increasing the viscoelasticity of samples).
This agrees with the experimental observation, (Fig.~\ref{master-exponent} a).
We should mention that the theoretical exponent just maps very first data points of experimental results ($\max(\eta) \rightarrow  $ \SI{10}{\milli \pascal \second}), see SI.

In summary, this simple perturbation analysis confirms the major tendencies observed in the experiments: 
i) Adding  viscoelasticity to the system in terms of a shear-rate dependent viscosity, the effective exponent ($\alpha^{'}$) decreases.
ii) Increasing the viscous dissipation (i.e., the perturbation term), the prefactor decreases. 
This simple proposed model shows the key features of the experimental tendencies.

\section{Conclusion}
The early stage spreading of Newtonian and viscoelastic fluids on hydrophilic and hydrophobic substrates has been studied. 
Generally speaking, viscoelastic drops spread faster compared to the Newtonian case with the same physical properties (zero shear rate viscosity and surface tension). 
This difference can be justified by the fact that near the contact line, the shear rate is extremely high. 
This leads to decreasing the effective viscosity which is not the case for Newtonian liquids. 
To be able to observe the viscoelastic effect, the experimental time scale should be in order of internal relaxation time of the used polymer solution or longer. 
These experimental observations can be supported by a simple perturbation model. 
The results also confirm the dependency of the spreading exponent to the wettability of the substrate.

\section {Funding}

This study was funded by the Deutsche Forschungsgemeinschaft Project No. 265191195–SFB 1194, “Interaction between Transport and Wetting Processes” and the Deutsche Forschungsgemeinschaft (DFG) Project No. 422852551,  within the priority program SPP 2171.

\bibliographystyle{jfm}
\bibliography{jfm}


\end{document}